# Comment to:

# "Particle–hole symmetry breaking in the pseudogap state of Bi2201"

Itai Panas

Department of Chemistry and Biotechnology

Chalmers University of Technology

S-412 96 Gothenburg, Sweden

**Abstract**

Shen et al. [1] recently reported on ARPES data from Pb-Bi2201 revealing both particle–hole symmetry breaking and pronounced spectral broadening, which they interpret to be indicative of spatial symmetry breaking without long-range order at the opening of the pseudogap. Here is demonstrated how their results could be interpreted to reflect static and dynamic inhomogeneous crystal fields causing inter-band transfer of holes upon cooling below T*. Possible relevance to formation of charge carrier inhomogeneities is discussed, and put in perspective of a proposed quantum chemical understanding of HTS.

**$Bi_{1.5}Pb_{0.5}Sr_{1.5}La_{0.5}CuO_6$ band structure deconvolution**

Causes for the observations reported for $Bi_{1.5}Pb_{0.55}Sr_{1.6}La_{0.4}CuO_{6+\delta}$ in [1] are sought in the $Bi_{1.5}Pb_{0.5}Sr_{1.5}La_{0.5}CuO_6$ model system by means of spin polarized GGA PBE band structure calculations. Taking the $Bi_2Sr_2CuO_6$ crystal structure as point of departure, in what follows the influences of different structural replacements of 25% $Sr^{2+}$ - $Bi^{3+}$ pairs by $La^{3+}$ - $Pb^{2+}$ pairs on the resulting electronic structures are demonstrated. In Figure 1, we note that there are three distinctly different positions for the 25% replacement of $Sr^{2+}$ by $La^{3+}$, and 25% replacement of $Bi^{3+}$ by $Pb^{2+}$. In all three cases, the ground state is a singlet. However, in all three cases a triplet state is only ~0.2 eV above the corresponding singlet. Figure 2 A, and B depict the spin densities of the triplet states corresponding to structure A and C in Figure 1. It is noted how the spin densities in the planes adjacent to $Sr^{2+}$-$La^{3+}$ ion pairs reflect a stronger crystal field than the planes exposed to $Sr^{2+}$-$Sr^{2+}$ ion pairs. Note in particular how the spin density on O in the $Sr^{2+}$-$Sr^{2+}$ bracketed planes acquire doughnut shape (superposition of $2p_\sigma$ and $2p_\pi$) where O in the $Sr^{2+}$-$La^{3+}$ bracketed planes display a dumbbell shaped spin density, i.e. $O2p_\sigma$, see Figure 2 again. The fact that in the present case crystal field inhomogeneities produce corresponding charge carrier segregations is difficult to arrive at simply by looking at band structures and corresponding densities of states DOS:s (Figure 3) of the singlet states. While some electronic reorganization can be noted upon increase La-Pb separation it is difficult to draw any detailed conclusions from the results presented in Figure 3. In order to make further connection to Figure 2, the triplet state α-spin Partial Densities of States for the $CuO_2$ planes are displayed subject to strong and weak crystal fields, respectively (see Figure 4). It is noted how for each structure the PDOS of each of the two planes is similar, while the Fermi level is displaced by the local crystal field (compare Figures 4 A&D, B&E, C&F) . Thus, the charge carrier inhomogeneities shown for the triplet states in Figure 2 are clearly reflected in the PDOS:s of the triplet states shown in Figure 4. The spin density cannot be employed to read whether any charge carrier inhomogeneity is present in the singlet state because the ground states does not display any net spin. However, having learned how said inhomogeneities are represented in the PDOS:s (compare Figures 2 and 4 again) in Figure 5 is plotted the PDOS:s for the structures in Figure 1 in their singlet states. Again, the PDOS:s in the vicinity of the Fermi level display clear similarities, and again it is observed how the Fermi level is "off-set" differently in planes experiencing the strong and weak crystal fields. This implies an electron transfer from the $CuO_2$ planes experiencing the weak crystal fields into the planes experiencing the strong crystal fields. Now, we may return to the band structure (Figure 3) and interpret it to reflect the changing separation between La-Pb. Moreover, by comparing the band structure in Figure 3 to that of the undoped $Bi_2Sr_2CuO_6$ (Figure 6A) it is clearly seen how bands in the vicinity of $(0,\pi)$ along the $(0,\pi)$-$(\pi,\pi)$ direction become shifted towards the Fermi level in case of the $La^{3+}$-$Pb^{2+}$ doped samples. In addition, the apparent single band which crosses the Fermi level from $(\pi,\pi)$-$(0,0)$ in Fig. 6A, is seen to split upon replacement of $Sr^{2+}$ - $Bi^{3+}$ pairs by $La^{3+}$ - $Pb^{2+}$ pairs (cf. Figure 3A). A third marked effect is seen by comparing the PDOS:s of the inequivalent $CuO_2$ planes in $Bi_{1.5}Pb_{0.5}Sr_{1.5}La_{0.5}CuO_6$ in the singlet state (Figure 5) to the corresponding PDOS in $Bi_2Sr_2CuO_6$ (Figure 6B). Note the pronounced double-peak feature at ~0.3 and ~0.7 eV in Figure 6B, and how it is off-set upwards in the weak field environments (Figure 5A-C), while in the strong-field environment it is down-shifted. Partial connection between the PDOS:s of the $Bi_2Sr_2CuO_6$ and $Bi_{1.5}Pb_{0.5}Sr_{1.5}La_{0.5}CuO_6$ compounds is made by displacing two $Sr^{2+}$ ions pairs 0.1 and 0.2 Å away from the bracketed $CuO_2$ plane, Figures 6C-D and Figures 6E-F, respectively. Thus is artificially created a weak-strong field inhomogeneity between $CuO_2$ planes. Interestingly, while the changes in the weak field $CuO_2$ plane PDOS are recognized (compare Figure 5A-C, with Figure 6D and 6F), no changes are observed in the strong field

CuO$_2$ PDOS (compare Figures 6B and 6H). This implies that redistribution of electrons among the O2p$_\sigma$ and O2p$_\pi$ bands can occur also in the absence of inter-plane charge transfer, by the displacement of the large cations in the vicinity of the CuO$_2$ planes. Note in Figure 6C (6E) how a band rises and crossed the Fermi level in the $(0,\pi)-(\pi,\pi)$ direction reflecting said redistribution of electrons (compare Figure 6A, 6C, and trends in band structures in Figure 3).

**Possible interpretation of experiment [1]**

The proposed understanding is summarized in Figure 7. The Fermi surface of the holes doped cuprates at T>T* is well understood in terms of a doped Hubbard-Mott insulator based on local Cu3d$^{9-\delta}$. Upon cooling Shen et al. report the development of band shoulders at $\sim(\pm 0.1, \pi)$ for T~T* and their saturation at $\sim(\pm 0.2, \pi)$ for T<<T*. Figure 7 illustrates how the dispersive Cu3d band is holes doped at elevated temperature. Upon cooling, localization destabilizes the Cu3d holes. Due to the Sr$^{2+}$-Sr$^{2+}$ versus La$^{3+}$-Sr$^{2+}$ asymmetry holes are accumulated at the former sites. This can be understood as an increased stabilization of the $Cu\ 3d_{x^2-y^2}$ - $O2p_\sigma$ band at the latter plane to the extent that a second purely oxygen associated band, i.e. the $O2p_\pi$ band at the Sr$^{2+}$-Sr$^{2+}$ bracketed site becomes electron donating. Compare Figures 2A, and B where the spin densities on the O ions in the weak field planes acquire a doughnut shape indicative of both 2p$_\sigma$ and 2p$_\pi$ characters. This is as opposed to the strong field planes which display only O2p$_\sigma$ character. Indeed, the particle-hole asymmetry reported by Shen et al. [1] implies that the increasing number of electrons residing in the band composed of atomic $Cu\ 3d_{x^2-y^2}$ - $O2p_\sigma$ states (red) bracketed by La$^{3+}$-Sr$^{2+}$ ions requires a second oxygen band as electron source. This by itself infers that a $O2p_\pi$ band becomes hole doped (Figure 7, white) because it is the only other oxygen related band besides the O2p$_\sigma$, and the latter is employed to propagate the super-exchange interaction between Cu3d$^9$ sites. The least stable $O2p_\pi$ band is the one which is bracketed by the Sr$^{2+}$-Sr$^{2+}$ pairs.

In the above described calculations as summarized in Figure 7, the impact of the proximity of La$^{3+}$ to a CuO$_2$ plane, as compared to a Sr$^{2+}$ ion, with regard to charge carrier distribution was demonstrated. Three-dimensional random distribution of La$^{3+}$ ions replaces the layer-by-layer charge carrier segregation by zero-dimensional such inhomogeneities, and preserves the effective periodicity of the crystal. The *a prori* symmetry broken crystal field may be insignificant at elevated temperatures due to the high mobility of the charge carriers at those temperatures but may become decisive below T*, and thus produce the observed temperature dependence of the ARPES signal [1].

Charge carrier segregation resulting in super-lattice formation in the *ab*-plane due to a possibly generic laterally inhomogeneous crystal field is displayed in Figure 8A, where the O2p$_\pi$ character is emphasized. Though no direct proof for such is provided in [1], the spectral broadening, taken by Shen et al. to reflect electronic inhomogeneities, supporting an interpretation along the line suggested in Figure 7 because (a) accumulation of charge carriers in the low-dispersive $O2p_\pi$ bands is expected to display instability towards holes clustering, and (b) 25% random replacement of Sr$^{2+}$ by La$^{3+}$ is expected to cause random charge carrier inhomogeneities due to the inhomogeneous crystal field. Having thus addressed the static case, "dynamic" such random variations in crystal field (c-axis Sr disorder in Bi$_2$Sr$_2$CuO$_6$) were demonstrated above to cause redistribution of electrons among $O2p_\sigma$ and $O2p_\pi$ associated bands. Connection between Figure 6, 7, Figure 8A and ARPES [1] is made in Figures 8B-F. Thus, in Figure 8B is the O2p$_\pi$ character of the band crossing the Fermi level displayed (*cf.* Figures 6E and 6F). In Figure 8C is said band (cf. Figure 6A and 6B) folded twice to satisfy the super-cell periodicity (*cf.* Fig.8A). In Figure 8D is reflected the redistribution of charge carriers among the two bands due to change in relative stability, and connection made with Figure 6E-F and Figure 7C. Figures 8E and 8 finally connect to Figure 7D and the possible interpretation of the ARPES experiment [1] is again arrived at.

**A two-gapped multi-band scenario for High Critical Temperature Superconductivity**

The scenario for HTS formulated by us [2] includes three steps. (i) At elevated temperatures mobile holes reside in the bands produced by the $Cu\ 3d_{x^2-y^2}$ - $O2p_\sigma$ states. (ii) Upon cooling the charge carriers become trapped, e.g. as Zhang-Rice singlets [3], or transferred into bands of $O2p_\pi$ character. Indeed, recovery of anti-ferromagnetic coupling [4] requires such a transfer. The opening of the pseudo-gap has two contributions, one is (ii.a) the development of AF coupling among $Cu3d^9$ sites, and the second is (ii.b) the complementary clustering of holes [2,5]. (iii) HTS emerges from a two-gaped "normal" state, such that resonant coupling of virtual holes clusters excitations and virtual magnons contribute to the correlated ground state [2,5,6]. Aspects of this understanding have been articulated in terms of a real-space analog [2,5] to the Bardeen Cooper Schrieffer theory, and in an equivalent two-component RVB-BEC formulation [6]. The latter implies that BEC among virtual holes cluster excitations is mediated by BEC of virtual magnons. Because the one provides the coupling required for the other to condense, the corresponding two signatures (superconductivity and spin-flip resonance [7]) appear at the same temperature, i.e. at $T_C$. In [5], the understanding developed for the cuprates was employed to formulate the superconductivity in FeSe, i.e. in terms of an analogous multi-band scenario to that developed for the cuprates [2,5,6]. While such complexity is generally accepted in case of the Fe-chalcogenides and Fe-pnictides, single-band scenarios still dominate in case of the cuprates. The report by Shen et al. may provide the first solid ARPES based evidence in favor of a multi-band mechanism promoted by segregation in case of the HTS cuprates.

**Computational details**

The band structure calculations employ the CASTEP [8] program package within the Material Studios framework [9]. The GGA PBE functional [10] was employed. Core electrons were described by ultra-soft pseudopotentials, O(6 electrons), Cu(11 electrons), Sr(10 electrons), La(11 electrons), Pb(14 electrons), Bi(5 electrons), employing a 340 eV cut-off energy. Summations over the Brillouin zone employed a $7 \times 7 \times 1$ Monkhorst-Pack grid[11].

# Figure Caption

**Figure 1.**

Crystal Structures of $Bi_{1.5}Pb_{0.5}Sr_{1.5}La_{0.5}CuO_6$, where the position of the $La^{3+}$ (light blue) with respect to the $Pb^{2+}$ (grey) differs.

**Figure 2.**

Spin densities corresponding to Figures 1A, and 1C. Note the doughnut shaped spin density on O reflecting both $2p_\sigma$ and $2p_\pi$ characters in the CuO2 planes experiencing weak crystal field. This is in contrast to the CuO2 planes experiencing the strong crustal fields where the spin densities on the oxygens take a $O2p_\sigma$ dumbbell shape. In addition, in Figure 2B is seen that the apical oxygens take on some hole character, which is absent in Figure 2A.

**Figure 3.**

Densities of States and in-plane band structure $(0,0)–(0,\pi)–(\pi,\pi)–(0,0)$ corresponding to Fig. 1A (top), Fig. 1B (Centre), and Fig.1C (bottom).

**Figure 4.**

Triplet state. α-spin PDOS in weak and strong ligand fields for the structures corresponding to Fig. 1A (top), Fig. 1B (Centre), and Fig.1C (bottom). Note how in the vicinity of the Fermi level the PDOS:s are similar in the A-D, B-E, and C-F pairs, while the Fermi level is off-set differently by the different local crystal fields. In case of A-D compare Figure 2A. In case of C-F, compare Figure 2B.

**Figure 5.**

Same as in Figure 4, but for the ground state singlet state. PDOS in weak and strong ligand fields for the structures corresponding to Fig. 1A (top), Fig. 1B (Centre), and Fig.1C (bottom). Note again how in the vicinity of the Fermi level the PDOS:s are similar in the A-D, B-E, and C-F pairs, while the Fermi level is off-set differently by the different local crystal fields..

**Figure 6**

(A) Band structure, and (B) $CuO_2$ PDOS for $Bi_2Sr_2CuO_6$ (compare Figure 3). Also displayed are the band structure (C) and PDOS (D) for $CuO_2$ planes in $Bi_2Sr_2CuO_6$ caused to experience artificially weak field (0.1 Å displaced $Sr^{2+}$ ions). (E) and (F) are same as (C) and (D) but with 0.2 Å $Sr^{2+}$ displacements. (G) is the same as in (E), while (H) is PDOS for the $CuO_2$ plane in the modified $Bi_2Sr_2CuO_6$ structure, which experiences an unchanged local field (see text).

**Figure 7**

(A) Section probed by ARPES [1].

(B) The $Cu\ 3d_{x^2-y^2}$ - $O2p_\sigma$ conduction band (red), and the $O2p_\pi$ band (white) at T>T*.

(C) Mean-field stabilization of the $Cu\ 3d_{x^2-y^2}$ - $O2p_\sigma$ band in the strong field $CuO_2$ plane at T<T* implies relative destabilization of the $Sr^{2+}$-$Sr^{2+}$ $O2p_\pi$ band (weak field $CuO_2$ plane II). Conservation of doped holes implies inter-band electron transfer into the $La^{3+}$-$Sr^{2+}$ $Cu\ 3d_{x^2-y^2}$ - $O2p_\sigma$ band in the strong field CuO2 plane (I) from the $Sr^{2+}$-$Sr^{2+}$ $O2p_\pi$ band (II).

(D) Resulting band structure. Holes and electrons at $|k|>|k_F+\kappa|$ are Cu 3d like and $O2p_\pi$ like, respectively. The resulting $O2p_\pi$ holes, caused by the inter-band charge transfer, appear in the vicinity of (0,±π) and (±π,0), where a $Cu\ 3d_{x^2-y^2}$ - $O2p_\sigma$ band is found burried deep below the Fermi level. The opening of the gap has the local super-exchange interaction $J_{local}$ as relevant energy scale (green). A second source of stabilization is the clustering instability [2,5,6] (see Figure 8A).

**Figure 8**

(A) Spin density inhomogenity in a Hg1201 (4 × 4 × 1) super-cell, due to c-axis displacement of central $Ba^{2+}$ cations. (B) Kohn-Sham state at the Fermi level (see Fig.6E and 6F). Note the exclusive $O2p_\pi$ character of the state. (C-F) Folding of surfacing band in (0,π)−(π,π) direction to make contact with the super-cell representation (Fig.8a). (C) T>T* (compare Fig.6C, and Fig.7B), (D) T<T* (compare Fig. 6E and Fig.7C). (E) and (F) make connection to ARPES [1] (compare Fig.7D).

**Figure 1**

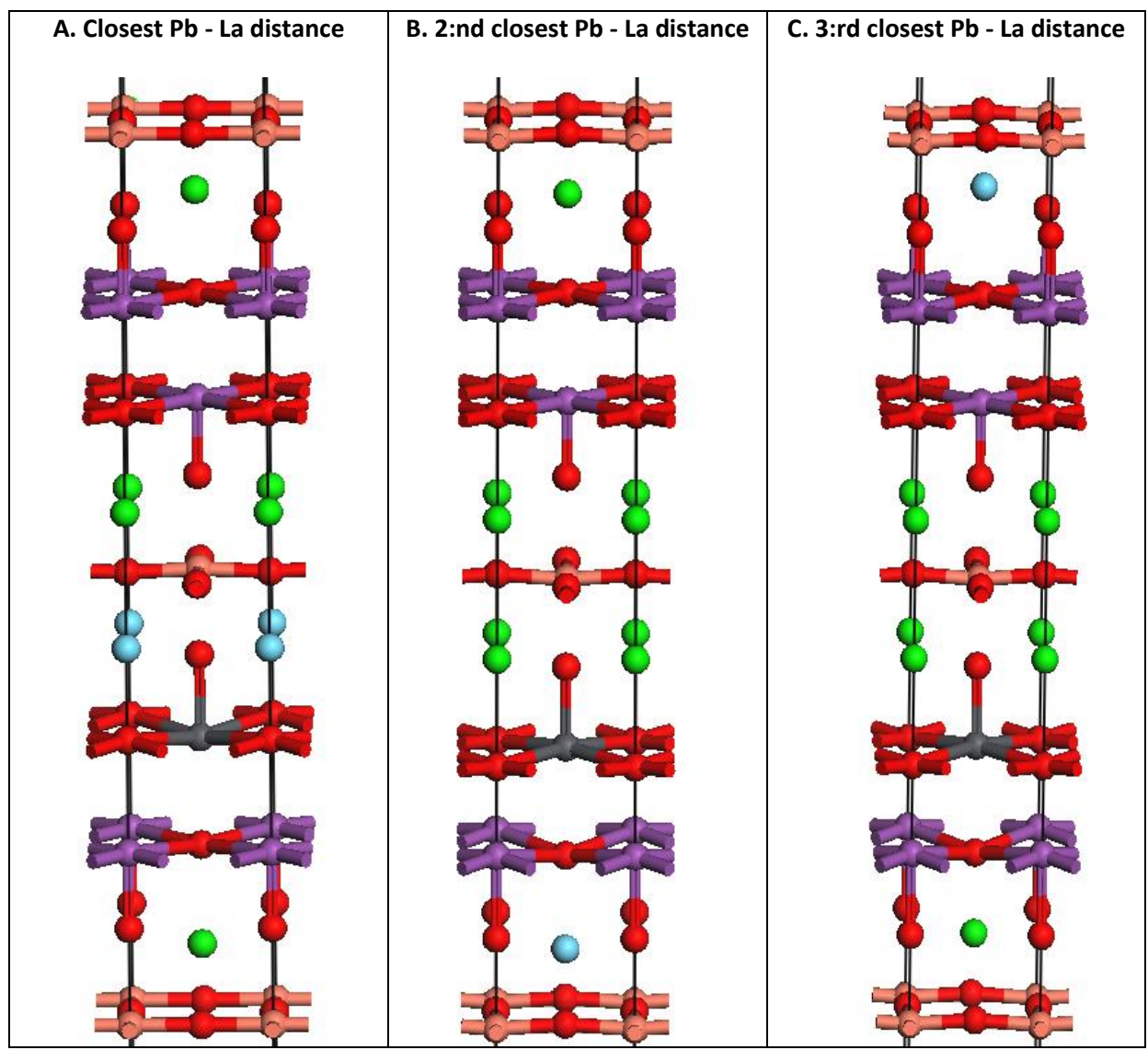

| A. Closest Pb - La distance | B. 2:nd closest Pb - La distance | C. 3:rd closest Pb - La distance |

**Figure 2.A.**

**Closest Pb - La distance**

**Weak field CuO$_2$ plane (top)**

**Strong field CuO$_2$ plane (bottom)**

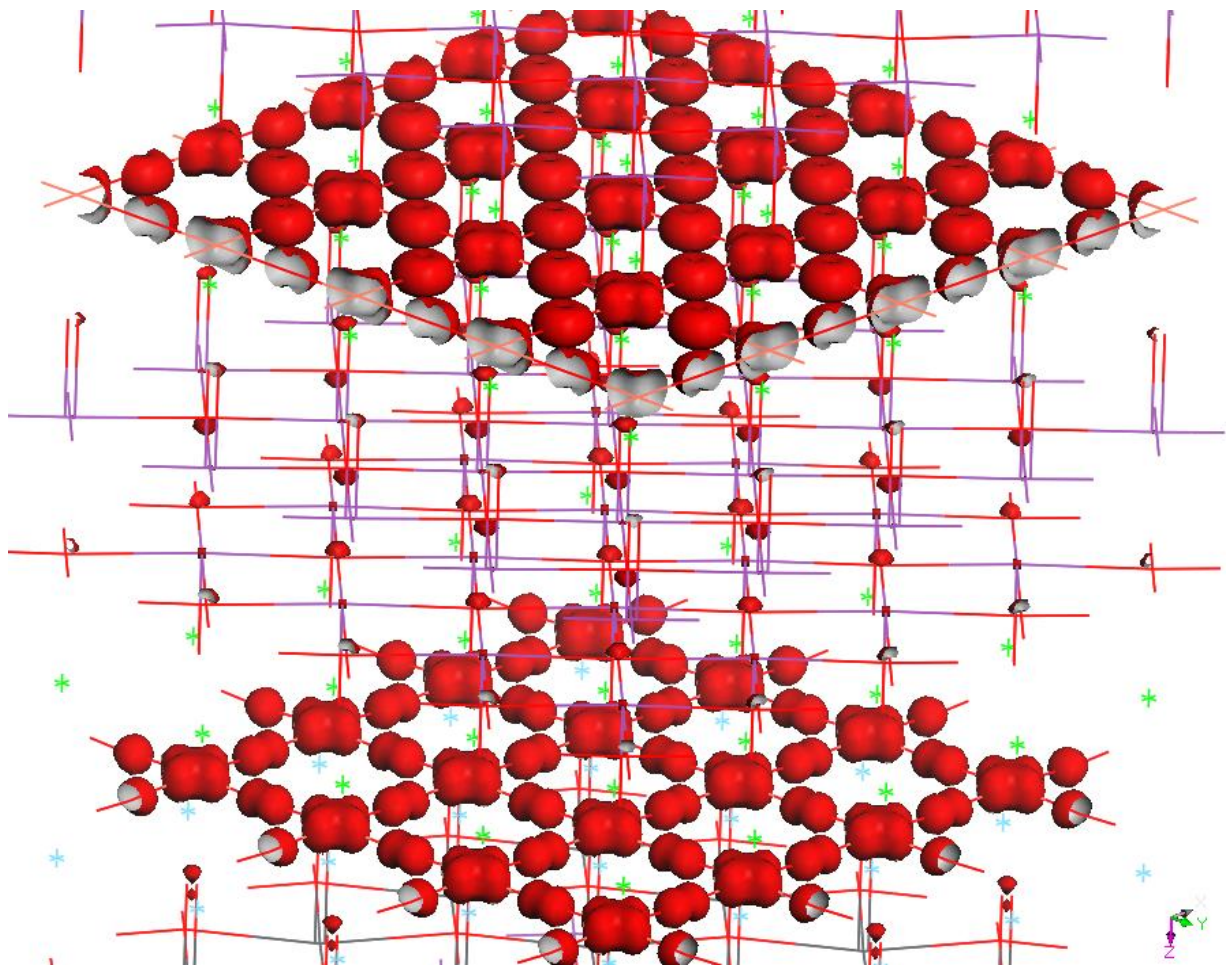

**Figure 2.B.**

**3:rd closest Pb - La distance**

**Weak field CuO$_2$ plane (top)**

**Strong field CuO$_2$ plane (bottom)**

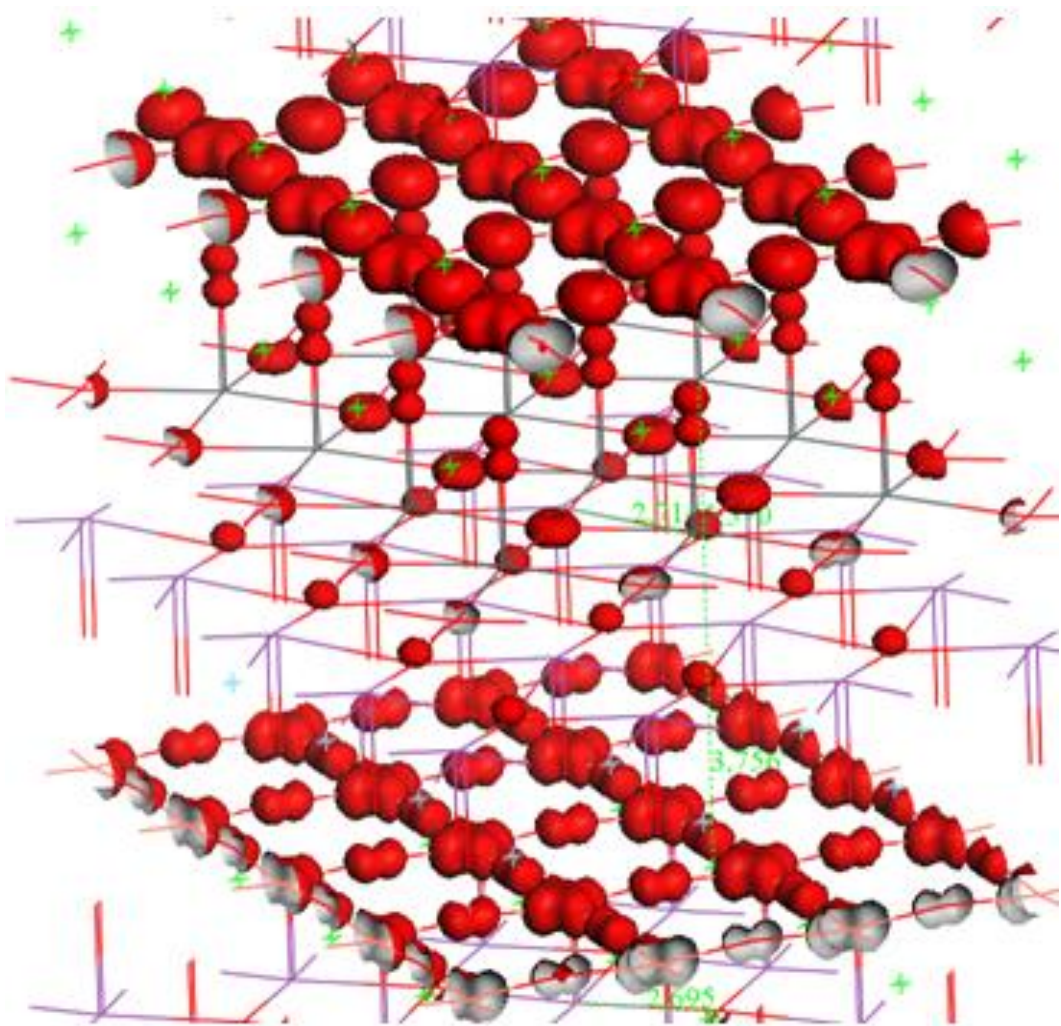

**Figure 3.**

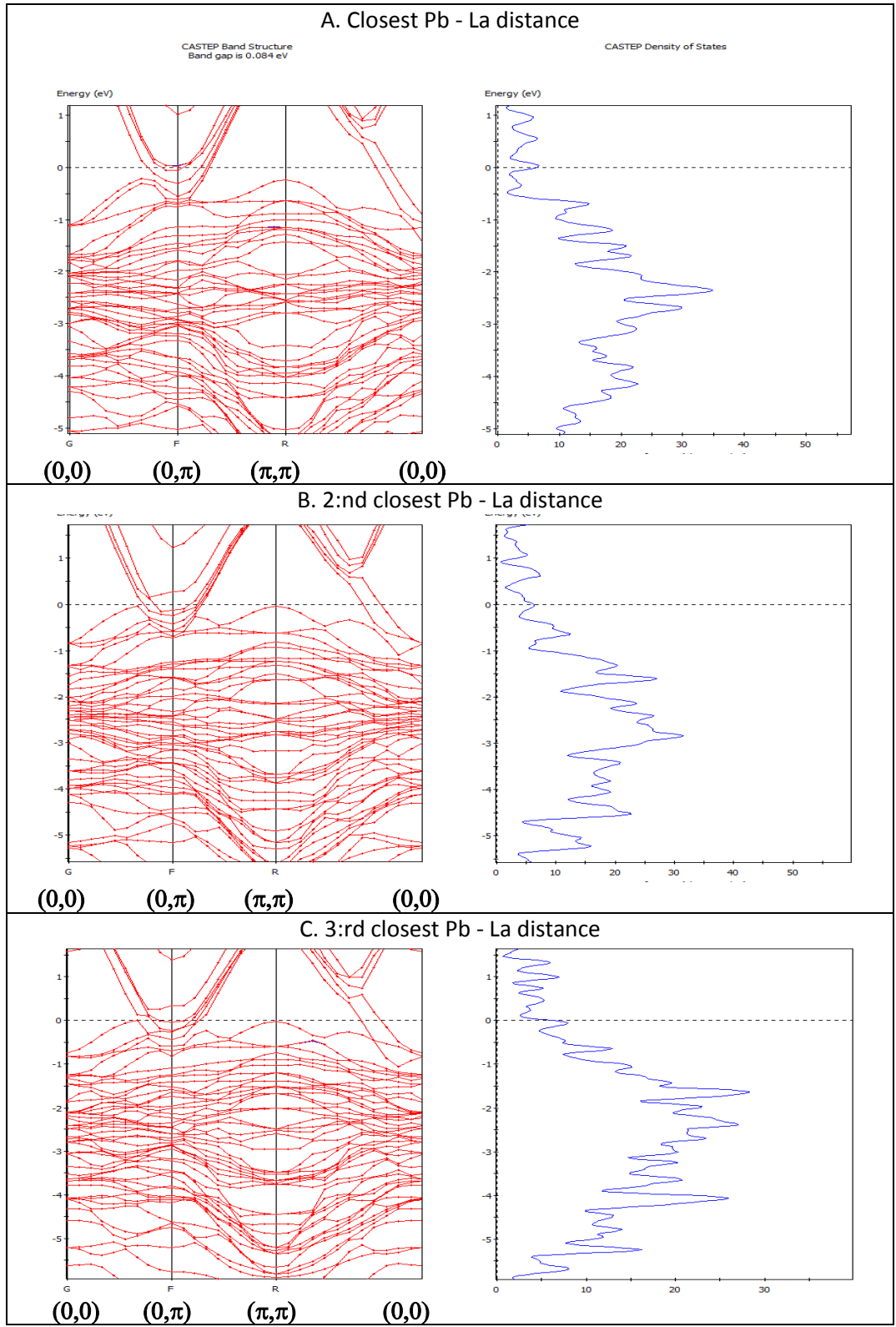

**Figure 4.**

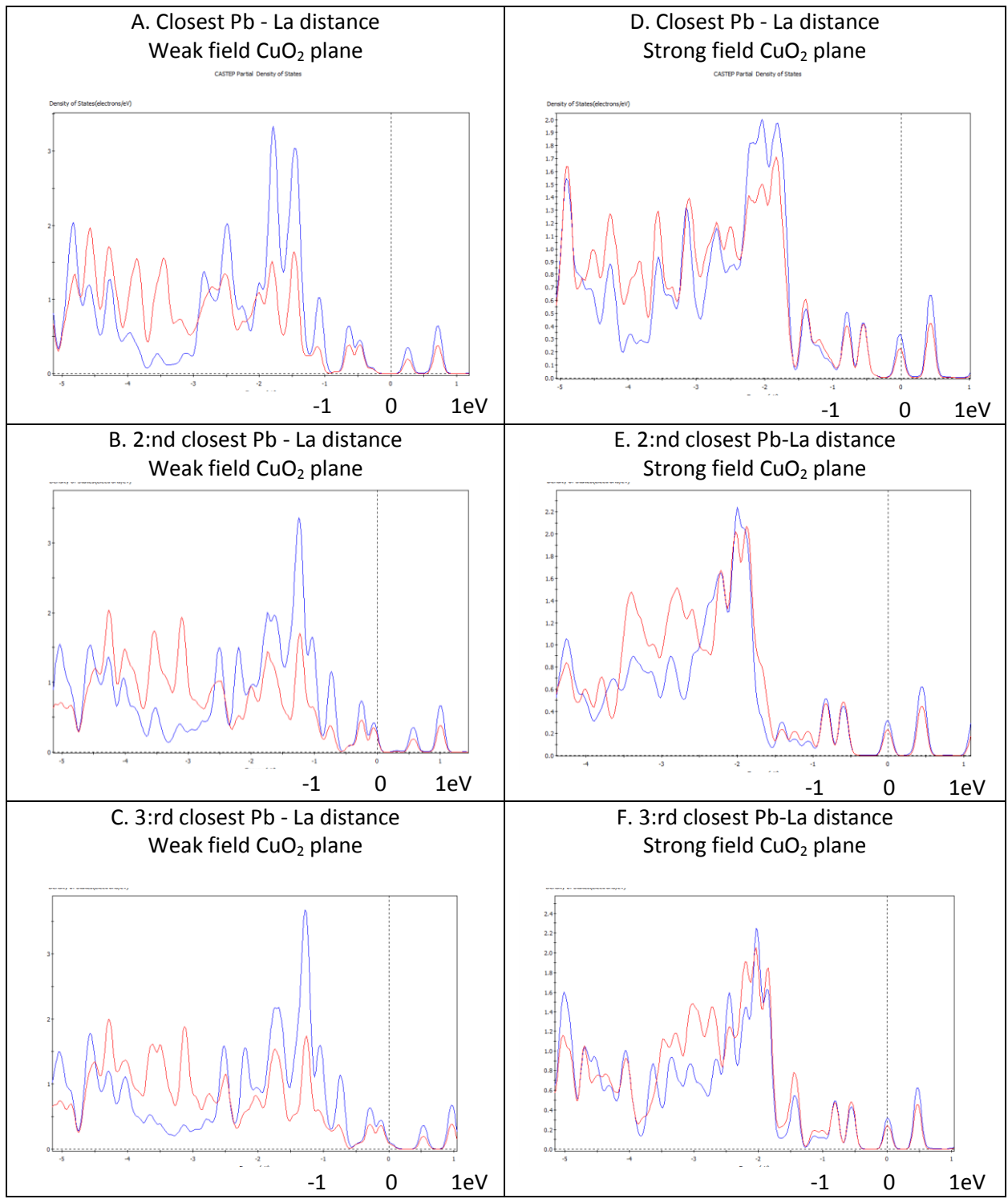

**Figure 5**

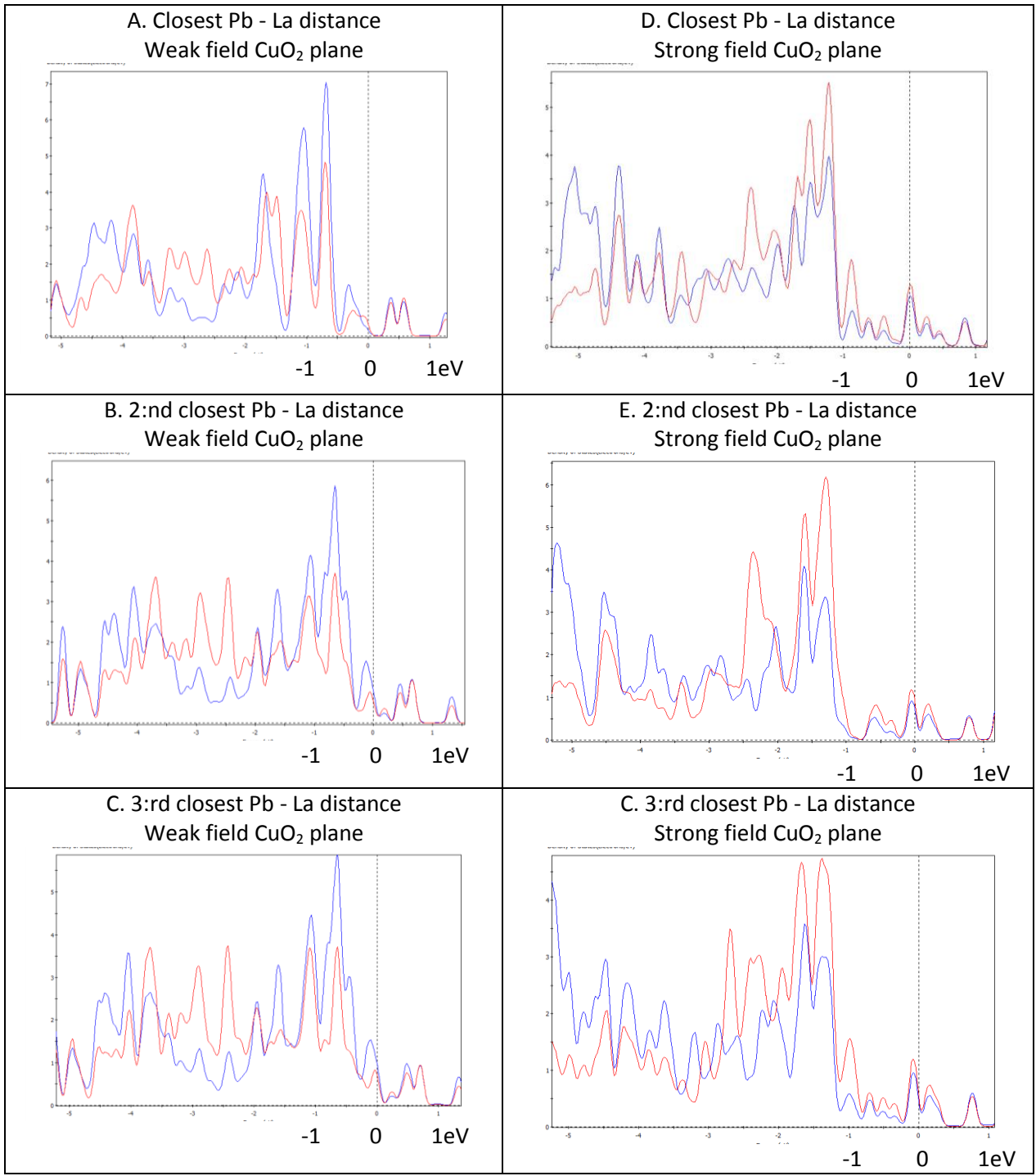

**Figure 6**

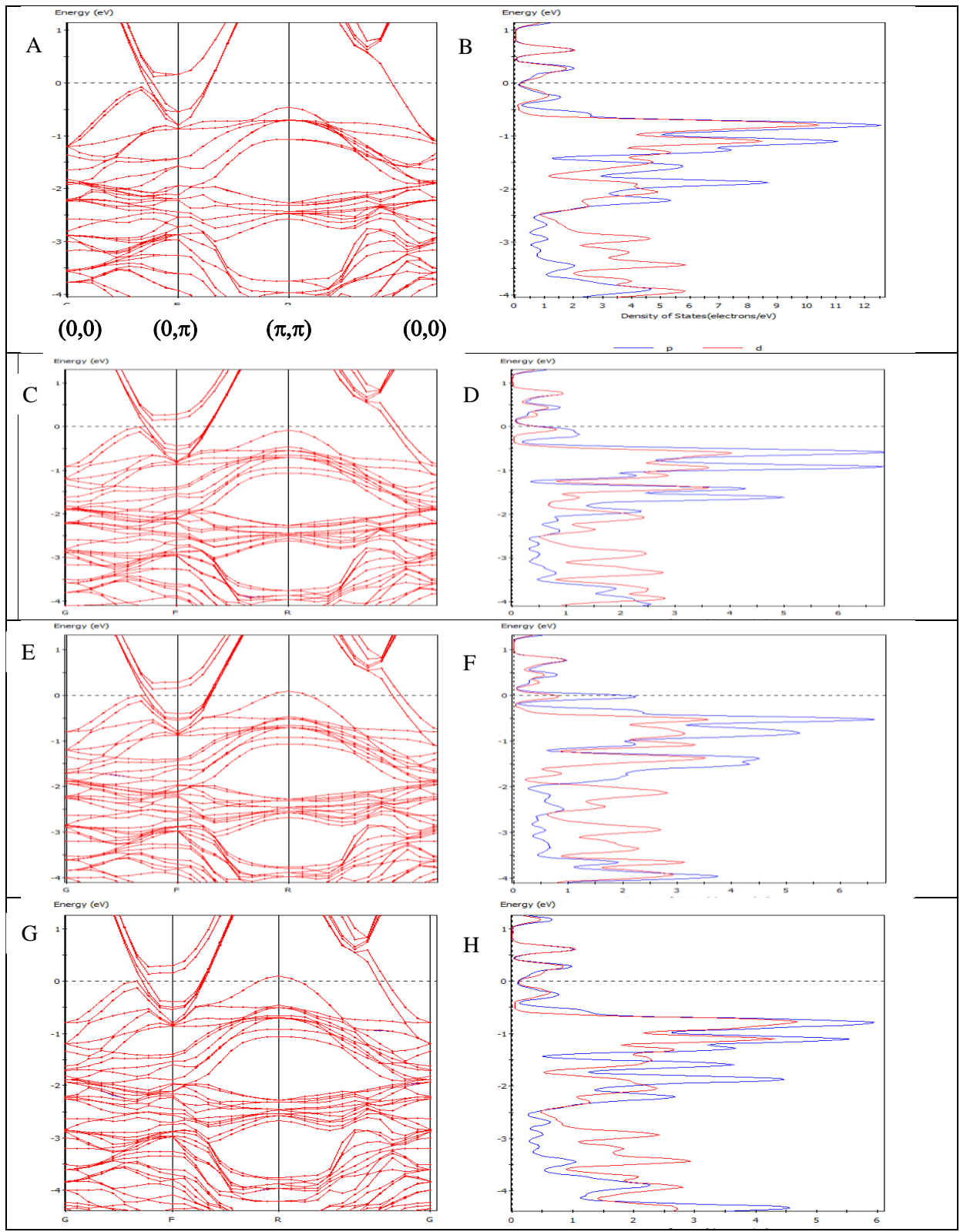

**Figure 7.**

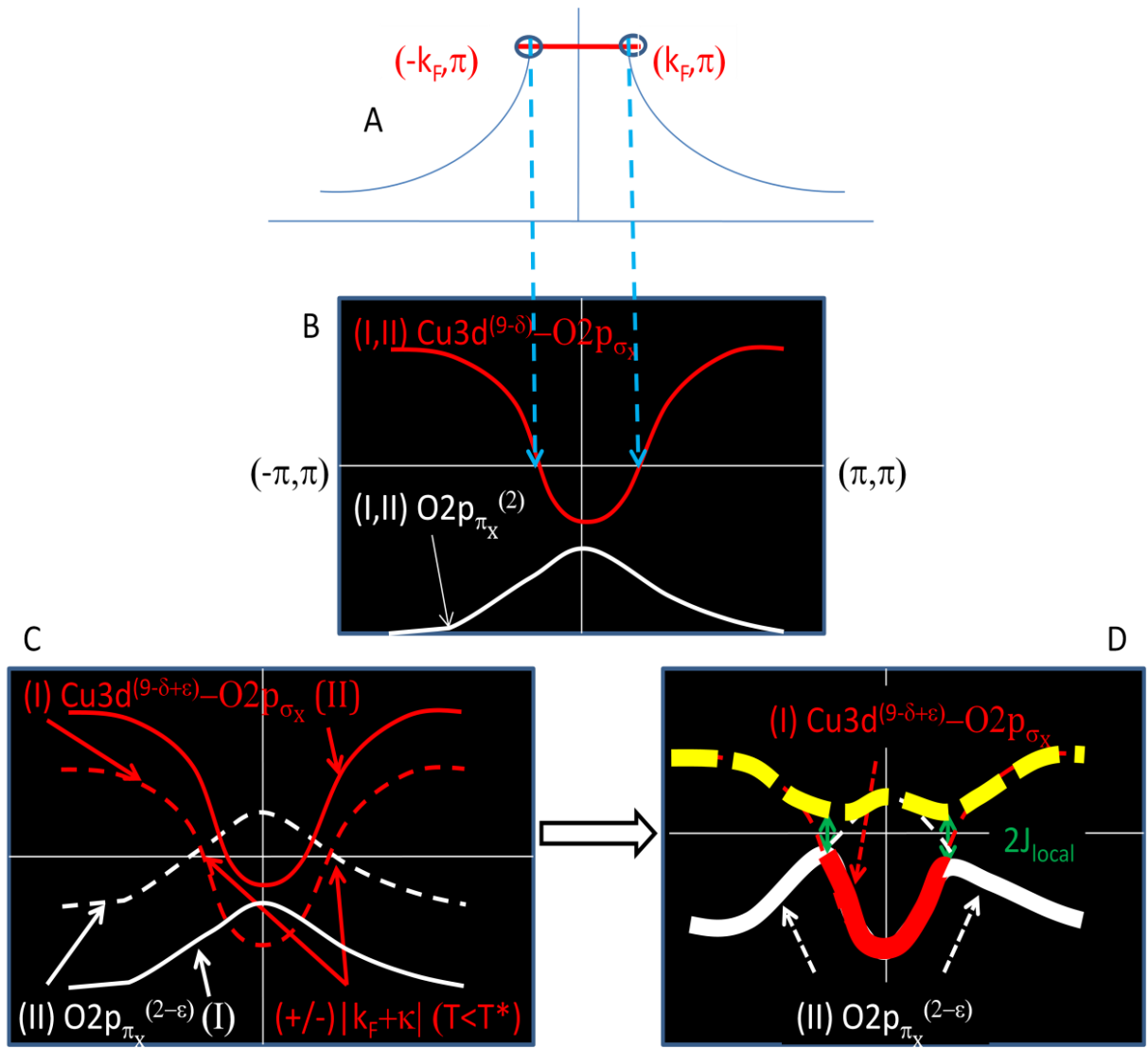

**Figure 8**

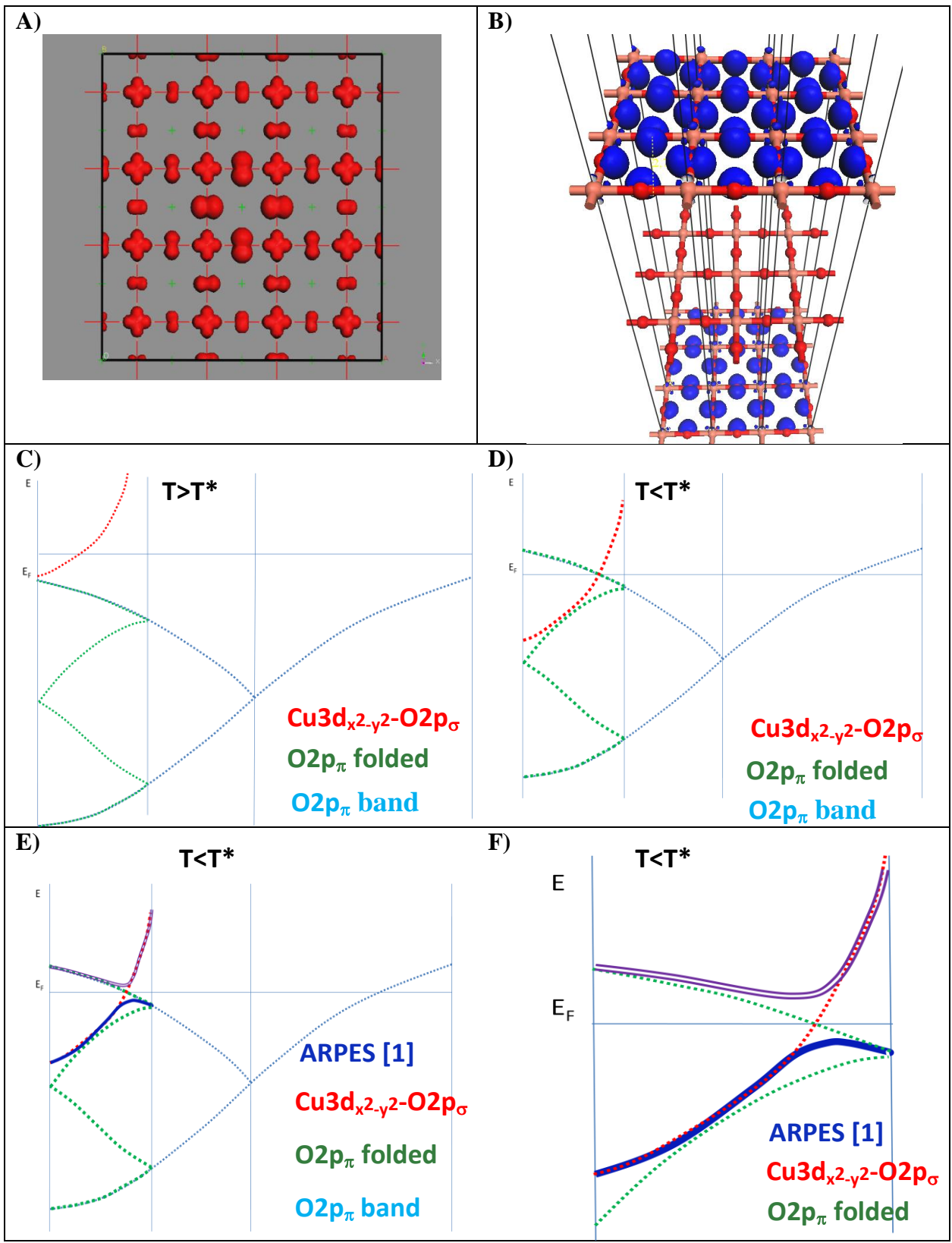